\documentclass[aps,prd,amsmath,amssymb,reprint,showpacs,superscriptaddress,nobibnotes,floatfix]{revtex4-1}
\usepackage{CDFRevTeX}

\usepackage{xspace}
\usepackage{standalone}
\usepackage{booktabs}
\usepackage{notoccite}
\usepackage{multirow}

\usepackage[tracking=true,kerning=true,expansion=true,spacing=false,factor=1100,stretch=20,shrink=20]{microtype}
\SetTracking{encoding={*}, shape=sc}{50}

\usepackage{siunitx}
\DeclareSIUnit\clight{\text{\ensuremath{c}}}

\usepackage[pdftex,unicode=true,pdfpagelabels=true,pdfusetitle,pdfauthor={The CDF Collaboration},hidelinks]{hyperref}

\usepackage{tikz}

\newcommand{\noteprefix}{CDF/ANAL/TOP/CDFR}
\newcommand{\notenumber}{11184}
\newcommand{\version}{3.2}

\begin{document}
\newcommand{\AFB}{\ensuremath{A_\text{FB}}\xspace}
\newcommand{\bbbar}{\ensuremath{b\bar{b}}\xspace}
\newcommand{\afbb}{\ensuremath{A_\text{FB}(\bbbar)}\xspace}
\newcommand{\gevM}[1]{\SI[per-mode=symbol]{#1}{\giga\eV\per\clight\squared}\xspace}
\newcommand{\gevMns}[1]{\SI[per-mode=symbol]{#1}{\giga\eV\per\clight\squared}}
\newcommand{\gevP}[1]{\SI[per-mode=symbol]{#1}{\giga\eV\per\clight}\xspace}
\newcommand{\gev}[1]{\SI[per-mode=symbol]{#1}{\giga\eV}\xspace}
\newcommand{\mbb}{\ensuremath{m_{\bbbar}}\xspace}
\newcommand{\dyb}{\ensuremath{\Delta y_b}\xspace}
\newcommand{\md}{\ensuremath{M_{\it{jj}}}\xspace}
\newcommand{\ttbar}{\ensuremath{t\bar{t}}\xspace}
\newcommand{\mvtx}{\ensuremath{M_{\text{vtx}}}}
\newcommand{\ptrel}{\ensuremath{p_{\mathrm{T, rel}}}}

\title{
Measurement of the forward-backward asymmetry in low-mass bottom-quark pairs produced in proton-antiproton collisions}

\date{\today}

\begin{abstract}
We report a measurement of the forward-backward asymmetry, \AFB{}, in $b\bar{b}$ pairs produced in proton-antiproton collisions and identified by muons from semileptonic $b$-hadron decays. The event sample is collected at a center-of-mass energy of $\sqrt{s}=$\SI{1.96}{\tera\eV} with the CDF II detector and corresponds to \SI{6.9}{\per\femto\barn} of integrated luminosity. We obtain an integrated asymmetry of \afbb{}$=(1.2 \pm 0.7)$\% at the particle level for $b$-quark pairs with invariant mass, \mbb{}, down to \gevM{40} and measure the dependence of \afbb{} on \mbb{}. The results are compatible with expectations from the standard model. 
\end{abstract}

\eprint{FERMILAB-PUB-16-017-E}


\maketitle

\section{Introduction}
\label{sec:Intro}
Asymmetries between distributions of heavy quarks and antiquarks produced in hadron collisions are important for tests of the standard model (SM) and searches for non-standard-model physics. The asymmetry arises from the interference of the different amplitudes contributing to the quark-pair production. Details of the origin of the asymmetries can be found, e.g., in Refs.~\cite{Rodrigo98, RodrigoKuhn, Bernreuter06}. An asymmetry in quark-pair production differing from the SM prediction is indicative of non-standard-model physics. 
Two asymmetries can be defined related to heavy-quark pair ($Q\overline{Q}$) production, the charge asymmetry, $A_\mathrm{C}$, and the forward-backward asymmetry, \AFB{}, as follows:
\begin{equation}
A_\mathrm{C}=\frac{N_{\mathrm{Q}}(\cos\Theta > 0)-N_{\mathrm{\overline{Q}}}(\cos\Theta > 0)}{N_{\mathrm{Q}}(\cos\Theta > 0)+N_{\mathrm{\overline{Q}}}(\cos\Theta > 0)}\;\mathrm{,}
\label{eq:AC}
\end{equation}
\begin{equation}
\AFB{}=\frac{N_{\mathrm{Q}}(\cos\Theta > 0)-N_{\mathrm{Q}}(\cos\Theta < 0)}{N_{\mathrm{Q}}(\cos\Theta > 0)+N_{\mathrm{Q}}(\cos\Theta < 0)}\;\mathrm{,}
\label{eq:AFBgen}
\end{equation}
\noindent
where $\Theta$ denotes the heavy quark $Q$ (anti-quark $\overline{Q}$) production angle in the incident parton-parton rest frame and $N_{\mathrm{Q}}$ ($N_{\mathrm{\overline{Q}}}$) is the number of quarks (anti-quarks) produced in the fiducial range. The parton momentum is taken as parallel to the incident proton, and similarly for the parton from the antiproton. 
Under the assumption of CP conservation in strong interactions, $N_{\mathrm{\overline{Q}}}(\cos\Theta > 0)=N_{\mathrm{Q}}(\cos\Theta < 0)$ and the two asymmetries are equal, $A_\mathrm{C} = \AFB{}$. 

Proton-antiproton ($p\bar{p}$) collisions at the Fermilab Tevatron with the center-of-mass energy of $\sqrt{s} = 1.96$~TeV allow for the study of asymmetries in pair-production of  of top-quarks ($t\bar{t}$) and bottom-quarks ($b\bar{b}$). The first measurement of the forward-backward asymmetry in \ttbar production that showed a difference between the measured \AFB value and the SM prediction was reported by the Collider Detector at Fermilab (CDF) experiment~\cite{afbtt2011}. Later, in other CDF measurement, the difference with a significance between 2 and 3 standard deviations, and increasing asymmetry with \ttbar invariant mass $m_{t\bar{t}}$, was reported~\cite{afbtt}. The discrepancy prompted significant theoretical activity, chiefly aimed at calculating predictions in various non-standard-model scenarios ~\cite{CaoRosner,*greshamkimzurek,*JungPearceWells,*BSMReviews,AguilarLightAxiG,*Schmaltz,*Falkowski,*Ipek}. Recent SM estimates of \AFB{} based on next-to-next-to-leading order (NNLO) quantum chromodynamic (QCD) calculations ease the discrepancy~\cite{Mitov,Kidonakis}. Additional measurements performed by D0~\cite{afbttd0} and CDF~\cite{afbttDil} are in good agreement with the higher-order calculations.
Similarly good agreement with the SM prediction is reported by the LHC experiments for the charge asymmetry in $t\bar{t}$ production~\cite{atlasAchBoosted, atlasAchLJ, atlasAchDil, cmsAch1, cmsAch2}.

Studying the \bbbar{} forward-backward asymmetry may provide additional constraints on non-standard-model physics scenarios. A CDF measurement of \AFB{} in \bbbar{} pair production at high \bbbar{} mass (\mbb{} $>$ \gevM{150}) gives results consistent with the SM predictions~\cite{afbbb}. The  measurement excludes a model with a \gevM{200} axigluon (gluon with axial coupling ~\cite{AguilarLightAxiG,*Schmaltz,*Falkowski,*Ipek}), whereas a model containing a heavier \gevM{345} axigluon is not excluded. The D0 collaboration has measured the \bbbar{} forward-backward asymmetry using reconstructed $B^{\pm}$ mesons~\cite{D0BmesonAfb}. The result is 3.3$\sigma$ below the SM prediction evaluated with {\sc{mc}}$@${\sc{nlo}}$+${\sc{herwig}}~\cite{mcnlo,herwig}, but consistent with zero asymmetry. In addition to the Tevatron measurements performed in $p\bar{p}$ collisions at $\sqrt{s}=$\SI{1.96}{\tera\eV}, the LHCb experiment has measured the \bbbar{}-production charge asymmetry in $pp$ collisions at $\sqrt{s}=$\SI{7}{\tera\eV}~\cite{LHCb}. The LHCb result is consistent with the SM expectations. 

The study presented in this article investigates \AFB{} in $p\bar{p}\rightarrow b\bar{b}X$ production down to \mbb{}$\,=\,$\gevM{40}. A muon from semileptonic $b$-hadron decays is used to distinguish between $b$ and $\bar{b}$ quarks. The fraction of \bbbar events in the data is estimated using a template fit based on the relative transverse momentum of the muon with respect to the axis of an associated jet~\cite{JetDef}, and the invariant mass of a secondary vertex within a second jet. The background-subtracted asymmetry is unfolded to particle level, where ``particle level'' refers to quantities reconstructed from final-state particles with lifetimes greater than $10\,$ps~\cite{partLevel}.  The article is structured as follows. Section~\ref{sec:Predict} briefly describes the origin of the asymmetry in heavy-quark pair production and the theoretical predictions. In Sec.~\ref{sec:Select}, the data sample and event selection are presented. The determination of the observed asymmetry from the data is described in Sec.~\ref{sec:Method}. The backgrounds are investigated in Sec.~\ref{sec:Bckg}. The unfolding procedure is discussed in Sec.~\ref{sec:unfold}. In Sec.~\ref{sec:Syst}, the systematic uncertainties are given.
The results are presented in Sec.~\ref{sec:Res}.

\section{\label{sec:Predict}Origin of production asymmetries and predictions}
In the SM, the two main strong-interaction pair-production processes are quark-antiquark annihilation ($q\bar{q}\rightarrow b\bar{b}$) and gluon fusion ($gg\rightarrow b\bar{b}$), neither of which induces an asymmetry at leading order (LO) in QCD. However, when higher-order corrections are considered, there are several sources of asymmetry~\cite{Rodrigo98, RodrigoKuhn, Bernreuter06}. Radiative corrections to quark-antiquark annihilation involve either virtual or real gluon emission, which leads to an asymmetry due to the interference of initial- and final-state radiative gluon processes (giving a negative contribution to \AFB{}) and interference of processes represented by so-called box and LO diagrams (giving a positive contribution to \AFB{})~\cite{Rodrigo98, RodrigoKuhn, Bernreuter06}. Interference of different amplitudes in flavor excitation of $q+g$ processes leads to an asymmetry, but the contribution is small with respect to the $q\bar{q}$ contribution~\cite{RodrigoKuhn, RodrigoKuhn2008}. Additional contributions to the asymmetry are expected from the interference with electroweak (EW) production processes mediated by $Z$ bosons or virtual photons, $q\bar{q}\rightarrow{Z/\gamma*}\rightarrow{b\bar{b}}$, which are at the level of $10$\%~\cite{RodrigoKuhn,GrinsteinMurphy, ManoharTrott}. No asymmetry is expected in $gg$ processes also at higher orders.

At the Tevatron, \bbbar production occurs predominantly through gluon-gluon fusion unlike top-quark-pair production where the dominant production process is $q\bar{q}$ annihilation.
As a consequence, when the full cross section is considered including contributions from $gg$, $q\bar{q}$, and $q(\bar{q})g$ interactions, the integrated asymmetry predicted by the SM is small. However, it is possible to enrich the sample in the $q\bar{q}\rightarrow b\bar{b}$ fraction with appropriate selection criteria, which can lead to a sizable forward-backward asymmetry.

There are several theoretical predictions predictions for \afbb that cover the low $b\bar{b}$ invariant mass region investigated in this study~\cite{RodrigoKuhn, GrinsteinMurphy, Murphy}. The prediction presented in Ref.~\cite{Murphy}, which also includes mixed EW-QCD corrections is summarized in Table \ref{Predict_tab}. Near the $Z$ pole, the SM bottom-quark asymmetry is maximal because it is dominated by tree-level exchange of EW gauge bosons. The selection criteria used in this analysis match the criteria used in~\cite{Murphy}, except for the transverse momentum requirement on the $b$ and $\bar{b}$ quark, $p_{\mathrm{T} b,\bar{b}}$: we require particle-level jets to have transverse energy $E_{\mathrm{T}} > 20\;$GeV~\cite{CoordinateSys}, while $p_{\mathrm{T} b,\bar{b}} > 15\;$GeV/$c$ is used in Ref.~\cite{Murphy}. The events with $p_{\mathrm{T} b,\bar{b}} < 20\;$GeV/$c$ can influence the $b\bar{b}$ asymmetry at low \mbb{}, that is only the first bin in Table \ref{Predict_tab}, for which the prediction has the highest uncertainty. 

\begin{table}[!h]
\caption{\AFB{} prediction from Ref.~\cite{Murphy} for different regions of $b\bar{b}$ invariant mass. The integral values for each bin are shown. The first contribution to the uncertainties on the predictions comes from neglecting higher-order QCD terms, while the second comes from varying the factorization and renormalization scales.}
\begin{tabular}{cc} \hline \hline
\mbb [\gevM{}] & \afbb [$\%$] \\ \hline 
35 -- 75  & $0.19 \pm 0.06\;^{-0.01}_{+0.01}$ \\
75 -- 95  & $2.49 \pm 0.16\;^{+0.52}_{-0.44}$ \\
95 -- 130 & $1.44 \pm 0.27\;^{+0.16}_{-0.12}$ \\
$>130$      & $2.14 \pm 0.63\;^{-0.01}_{+0.03}$ \\ 
Inclusive   & $0.34 \pm 0.08\;^{+0.01}_{-0.00}$ \\ \hline \hline
\end{tabular}
\label{Predict_tab}
\end{table}

\section{\label{sec:Select}The CDF II Detector and event selection}
The CDF II detector~\cite{Acosta:2004yw} is a forward-backward and cylindrically symmetric detector designed to study $p\bar{p}$ collisions with a center-of-mass energy $\sqrt{s}=$\SI{1.96}{\tera\eV} at the Fermilab Tevatron collider. The detector is approximately hermetic over the full angular coverage and is composed of a series of detectors to determine trajectories of charged particles embedded in an axial magnetic field of 1.4 T, surrounded by electromagnetic and hadronic calorimeters and muon detectors.

The analysis relies on the full data set collected by the CDF II detector of $p\bar{p}$ collision data but application of a prescaled online event-selection system (trigger) leads to the reduction of the integrated luminosity of the sample to \SI{6.9}{\per\femto\barn}. The process $p\bar{p} \rightarrow b\bar{b}X$ is analyzed using the so-called soft-muon technique, i.e., using the muon produced in the semileptonic decay of a $b$-quark to distinguish between $b$-jets from $b$ and $\bar{b}$-quarks. The trigger requires a muon candidate with transverse momentum p$_T>\;$\gevP{8} and pseudorapidity $|\eta| < 0.6$~\cite{CoordinateSys}. The offline selection requires at least two jets with transverse energies $E_\mathrm{T}>\;$\gev{20} and a muon candidate with $p_{T}>\;$\gevP{10} inside the cone of one of the jets (muon jet). The other jet (away jet) is required to be azimuthally opposed ($|\Delta\phi|>2.8$) with $|\eta|<1.0$ and balanced in $p_{T}$ with the muon jet. The $p_{T}$ balance condition requires that the difference between transverse momenta of the two jets does not exceed $60\%$ of the highest of the two. In addition, both the away and muon jets are identified as $b$ jets using two configurations of the secondary-decay-finding algorithm {\sc{secvtx}}~\cite{secVtxRef} and applying the more efficient configuration to the away jet. The algorithm identifies jets that most likely originate from the fragmentation of a $b$ quark by requiring the presence of charged-particle trajectories (tracks) that form reconstructable vertices significantly displaced from the vertex of the $p\bar{p}$ collision. The reconstructed jet $E_T$~\cite{CoordinateSys,JES} is corrected for the effects of jet fragmentation, calorimeter non-uniformities, and multiparticle interactions. Hence, the fiducial region of the measurement is defined by the following requirements: two azimuthally opposed () $b$-jets, balanced in $p_T$, with $|\eta| < 1.0$ and $p_{T}>\;$\gevP{20}; muon with $p_{T}>\;$\gevP{10} and $|\eta| < 0.6$ inside the cone of one of the $b$-jets.

The simulations use a {\sc{pythia}} (version 6.2.16~\cite{Pythia}) LO Monte Carlo dijet sample enriched in heavy flavor by requiring a muon with $p_{T}>\;$\gevP{8} and $|\eta|<1.2$. The \mbb{} distribution in events where a $Z$ or a virtual photon are produced is modeled by reweighting events from the {\sc{pythia}} Monte Carlo sample using the ratio of the LO differential cross sections of the QCD and EW processes computed using {\sc madgraph}~\cite{madgraph} as reported in Ref.~\cite{murphy_private}. A $10\%$ asymmetry~\cite{AfbZpole} is incorporated into the model of $Z-\gamma^{*}$ production.

For simulated events, the muon and away jets at particle level are defined in the same way as at the reconstructed level for real data. In addition we require matching of the muon and away jets at particle level with opposite-sign $b$ quarks using truth information from simulation. 

\section{\label{sec:Method}Asymmetry measurement}

The integrated forward-backward asymmetry for \bbbar pair production can be expressed using the difference of rapidities of the $b$ and $\bar{b}$ quarks, $\Delta y_b$~\cite{rapidity}, which is invariant under Lorentz transformations along the beam axis. For a $b$ quark from the \bbbar pair moving in the forward direction $\Delta y_b > 0$, and for the backward direction $\Delta y_b < 0$. The asymmetry \AFB{} in terms of $\Delta y_b$ is defined as follows:

\begin{equation}
\AFB{}=\frac{N(\Delta{y_{b}}>0)-N(\Delta{y_{b}}<0)}{N(\Delta{y_{b}}>0)+N(\Delta{y_{b}}<0)}\;\mathrm{.}
\label{eq:AFB}
\end{equation}
\noindent
In dijet events, where one of the jets contains a muon, $\Delta{y_b}$ is defined as follows:
\begin{equation}
\Delta{y_{b}} = Q(\mu)(y_{AJ}-y_{\mu J})\;\mathrm{,}
\label{eq:dYb}
\end{equation}
where $Q(\mu)$ is the charge of the muon, $y_{AJ}$ is the rapidity of the away jet, and $y_{\mu J}$ is the rapidity of the muon jet. Note that $\Delta y_b$ is positive if a $b$-quark is accompanied by a forward muon jet or a $\bar{b}$-quark by a backward muon jet, i.e., $\Delta y_b > 0$ if the $b$ quark is in forward direction and consequently $\Delta y_b < 0$ if the $b$ quark is in backward direction.

\begin{figure}[h!]
\includegraphics[width=\linewidth{}]{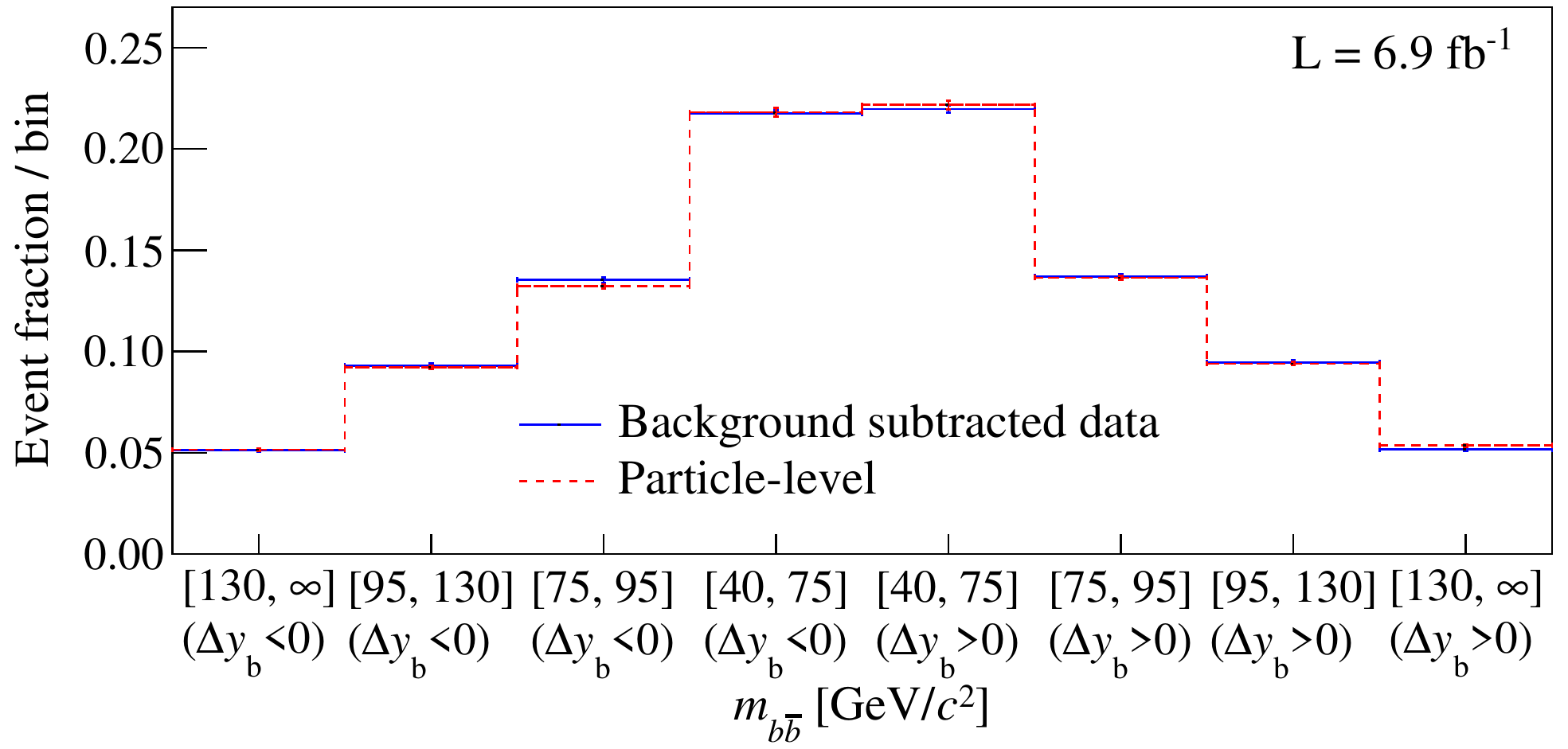}
\caption{Distribution of background-subtracted observed events prior to the unfolding (solid) and resulting particle-level events after unfolding (dashed) as a function of the combination of $b\bar{b}$ invariant mass and rapidity difference.}
\label{AfbInput}
\end{figure}

An unfolding procedure (see Sec.~\ref{sec:unfold}) is used to remove detector effects and infer \AFB{} in $b\bar{b}$ production at the particle level. To retrieve the particle-level \AFB{}, the background is subtracted from the observed distributions, and
the CDF II detector acceptance and resolution are taken into account. As the sign of $\Delta{y_b}$ depends on the charge of the muon, corrections for cascade decays $b \rightarrow c \rightarrow \mu$ and $B^0_{(s)}-\overline{B}_{(s)}^0$ mixing are included in the unfolding procedure. To measure \AFB{} at the particle level as a function of \mbb, the \mbb and \dyb distributions are simultaneously unfolded. We define a one-dimensional distribution of events divided into eight bins: two \dyb-bins (positive and negative \dyb), and four \mbb-bins ($[40, 75]$, $[75, 95]$, $[95, 130]$, and $[130,\infty ]\,$\gevMns{}) as shown in Fig.~\ref{AfbInput}, and perform a one-dimensional unfolding where $\infty$ stands for the kinematic maximum. 

\begin{figure*}[htbp!]
\centering
  \includegraphics[width=0.95\linewidth]{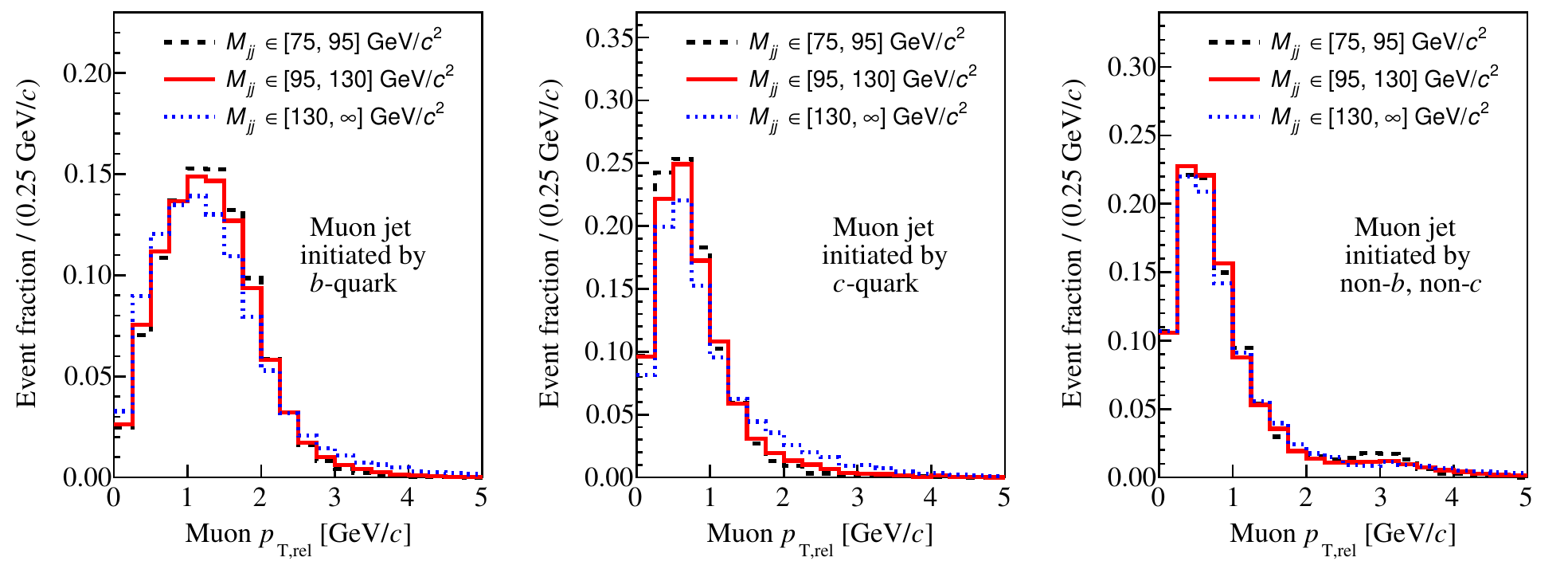}
  \caption{Distributions of \ptrel{} for simulated $b$-tagged jets on the muon-jet side, initiated by a $b$ quark (left), a $c$ quark (center), and a light quark or a gluon (right), which are used as templates in the fit of the signal fraction. Dashed, solid, and dotted lines correspond to different ranges in dijet mass.}
  \label{ptTemplate}
\end{figure*}

\begin{figure*}[htbp!]
\centering
  \includegraphics[width=0.95\linewidth]{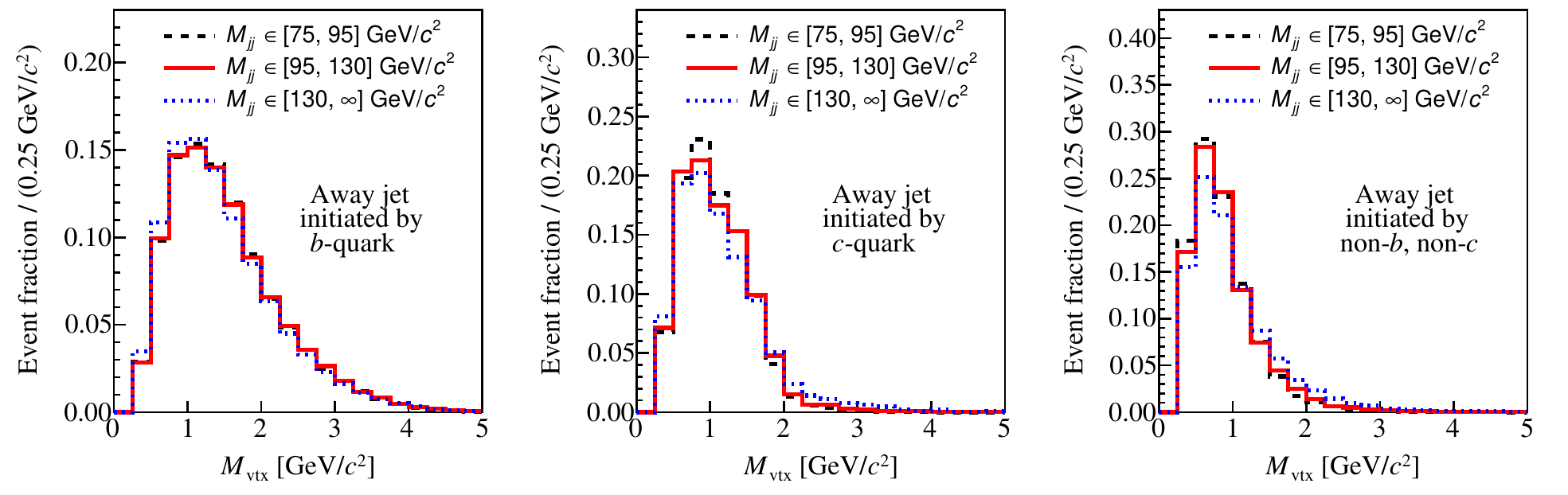}
  \caption{Distributions of \mvtx{} for simulated $b$-tagged jets on the away-jet side, initiated by $b$ quark (left), $c$ quark (center), and light quark or gluon (right), which are used as templates in the fit of the signal fraction. Dashed, solid, and dotted lines correspond to different ranges in dijet mass.}
  \label{mvtxTemplate}
\end{figure*}

\begin{figure*}[htbp!]
\centering
\begin{minipage}{.47\textwidth}
  \centering
  \includegraphics[width=\linewidth]{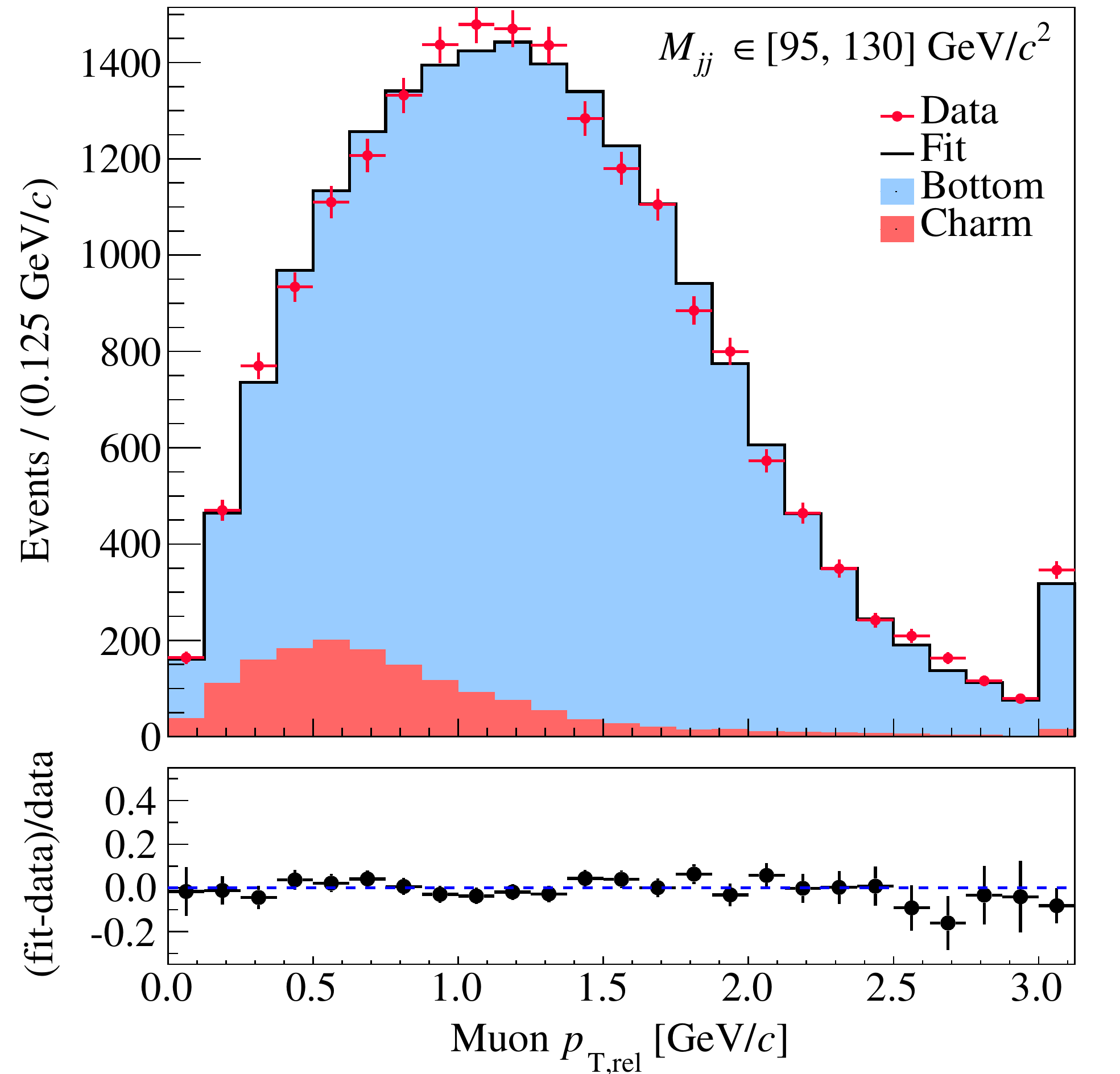}
  \caption{Distribution of \ptrel{} with two-component fit results overlaid for events restricted in \md{} of $[95, 130]\,$\gevMns{}. The last bin contains overflow entries. The lower panel compares the fit result with the data.}
  \label{PtrelFit}
\end{minipage}%
\hspace{0.5cm}
\begin{minipage}{.47\textwidth}
  \centering
  \includegraphics[width=\linewidth]{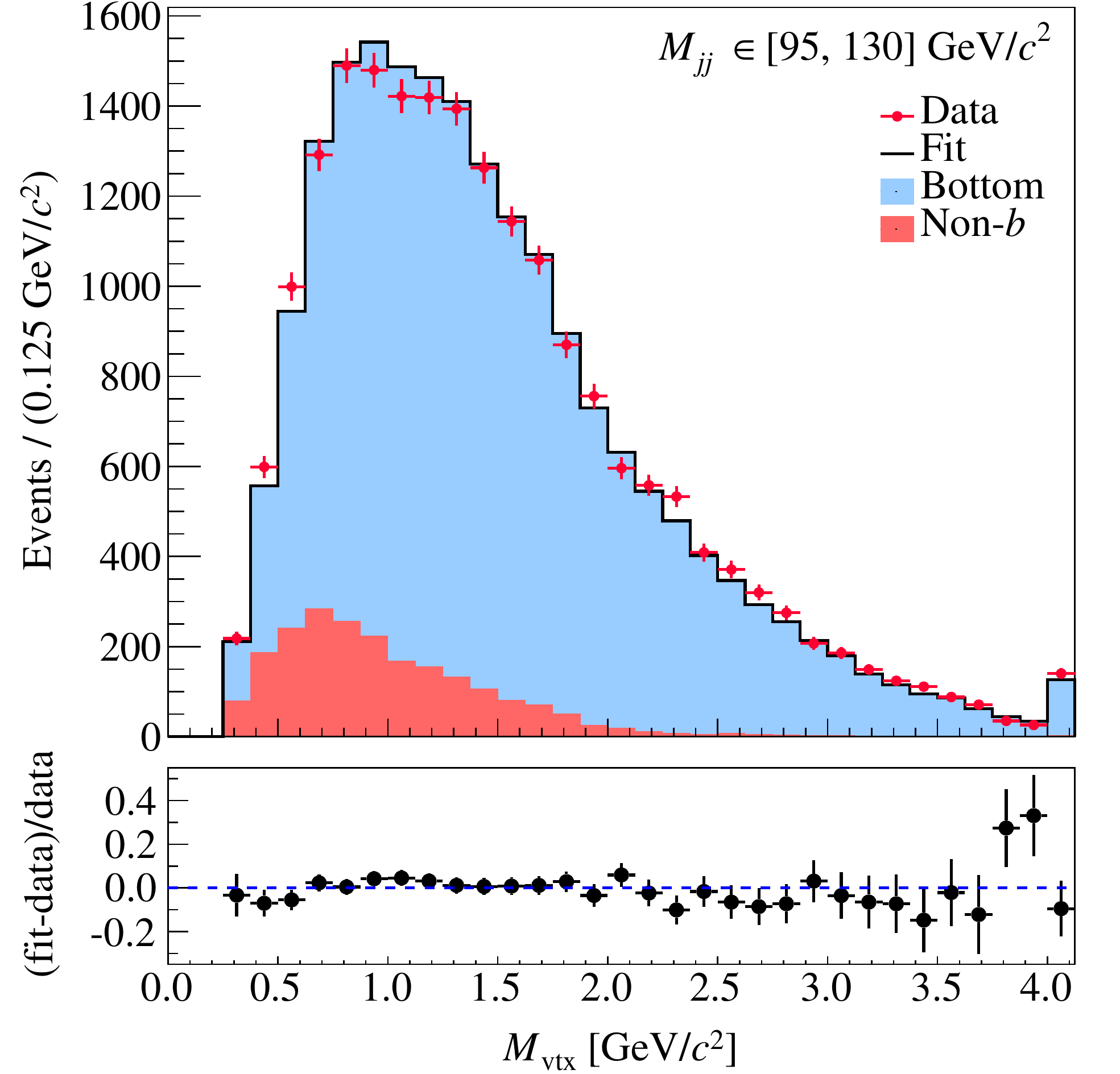}
  \caption{Distribution of \mvtx{} with two-component fit results overlaid for events restricted in \md{} of $[95, 130]\,$\gevMns{}. The last bin contains overflow entries. The lower panel compares the fit result with the data.}
  \label{MvtxFit}
\end{minipage}
\end{figure*}

To subtract the background, the fraction of \bbbar events in data, $f_{b\bar{b}}$, is estimated in four bins of the reconstructed dijet mass of muon jet and away jet, \md{}. The background distribution is then obtained by applying the weighting factor $1-f_{b\bar{b}}$ to the events in each \md{} bin. The fraction $f_{b\bar{b}}$ is obtained by determining the $b$ fractions in muon and away jets using two independent template fits. To extract the $b$ content of the muon jet, $f_{b}^{\mu J}$, we use the distribution of the component of the momentum of the muon perpendicular to the jet axis, \ptrel{}, which tends to peak at larger values for a $b$-quark jet than for a $c$-quark or light-quark jet as shown in Fig.~\ref{ptTemplate}. The templates for the $c$- and light-quark jets are nearly indistinguishable. Thus we consider only two templates in the fit, $b$ and $c$ templates, to obtain the fraction of events with the muon arising from a $b$-quark jet. To extract the $b$ content of the away jet, $f_{b}^{AJ}$, we perform a two-template fit to the distribution of secondary vertex mass, \mvtx~\cite{secVtxRef}, of the away jet, using mass templates for $b$- and non-$b$-quark jets. The template for the non-$b$-quark jets is created by merging the templates for the $c$- and light-quark jets, which are quite similar. The peak of the \mvtx{} distribution is correlated with the mass of the parton initiating the jet as shown in Fig.~\ref{mvtxTemplate}. Figures~\ref{PtrelFit} and~\ref{MvtxFit} show examples of the fits of the \ptrel{} and \mvtx{} distributions, respectively, for the \md{} bin $[95, 130]\,$\gevMns{}.

\begin{figure}[htbp!]
  \includegraphics[width=\linewidth{}]{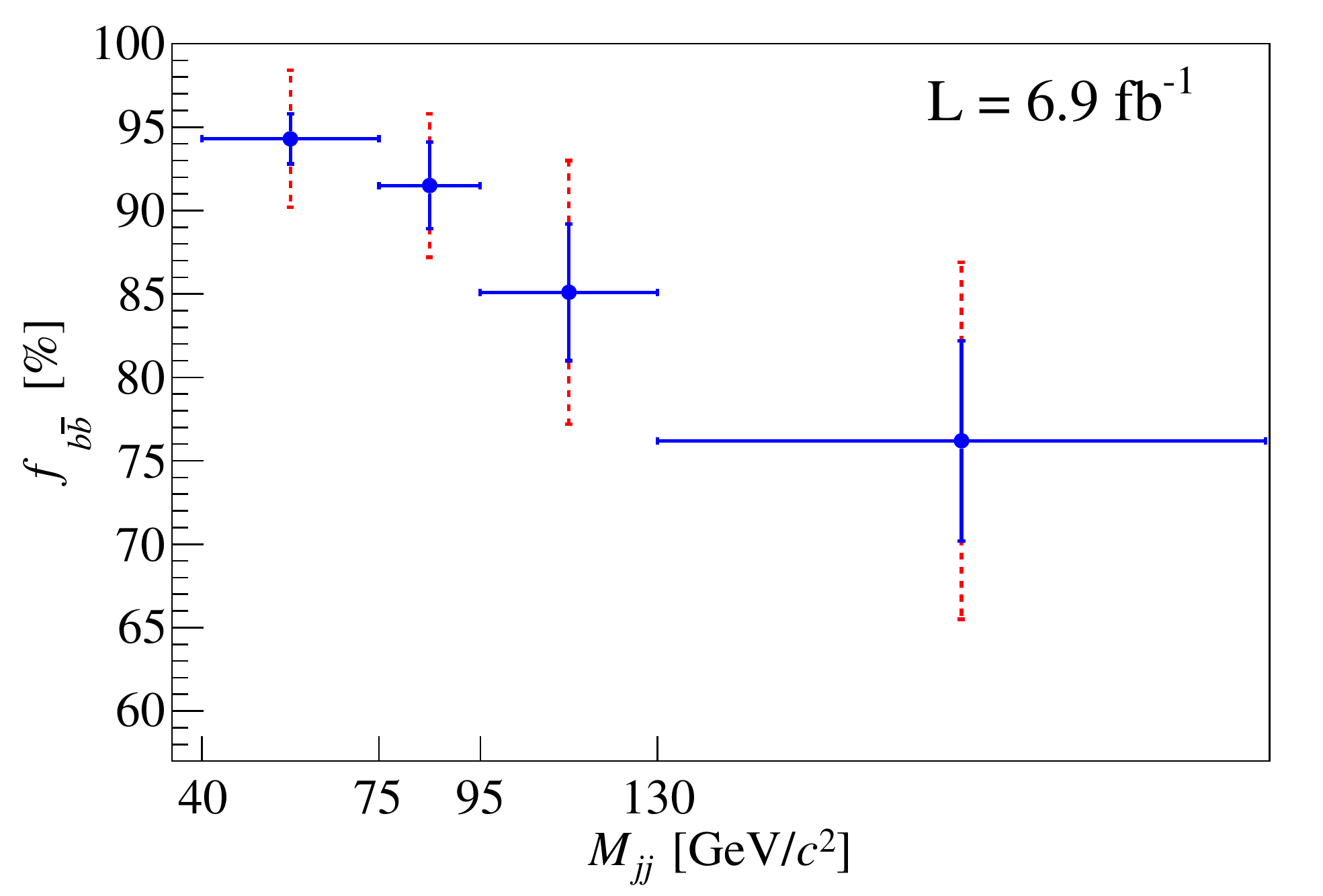}
  \caption{Distribution of $b\bar{b}$ fraction observed in data as a function of dijet invariant mass \md{}. The total uncertainties (outer bars) include the statistical (inner bars) and systematic uncertainties.}
  \label{fig:fbb_data}
\end{figure}

Since the $b$ fractions of the muon and away jets are obtained from independent fits, we have no information on their correlation in the dijet sample. However, if $f_{b}^{\mu J}>f_{b}^{AJ}$ we can obtain the highest (lowest) value of the \bbbar{} fraction by assuming that all non-$b$-quark muon jets corresponds to non-$b$-quark ($b$-quark) away jets, i.e., $f_{b\bar{b}}^{max} = f_{b}^{AJ}$ and $f_{b\bar{b}}^{min} = f_{b}^{AJ}-(1-f_{b}^{\mu J})$. We apply an analogous estimate to the case when $f_{b}^{\mu J}<f_{b}^{AJ}$. We then estimate the \bbbar{} fraction in each \md bin as the average of the upper and lower extremes in the bin, and set the corresponding uncertainty to the semidifference between the extremes. The results are shown in Fig.~\ref{fig:fbb_data}. The systematic uncertainties related to the procedure used to determine $f_{b\bar{b}}$ in data, coming from the fit strategy and template shapes, are summarized in Table~\ref{fbbsyst}.

\begin{table}[!h]
 \caption{Systematic and statistical uncertainties related to the procedure used to determine the $b\bar{b}$ fraction in data.}
 \begin{tabular}{lS[table-format=2.1]S[table-format=2.1]S[table-format=2.1]S[table-format=2.1]}
   \hline \hline
   \multicolumn{5}{c}{Absolute uncertainty of $f_{b\bar{b}}\,[\si{\%}]$} \\
   \multirow{2}{*}{Source} & \multicolumn{4}{c}{$\md\,[\si{GeV/\it{c}^2}]$} \\
   & {40 -- 75 } & { 75 -- 95 } & { 95 -- 130 } & { $> 130$} \\
   \hline
   {Fit strategy}         &1.6 &0.8 &2.2 &1.3  \\
   {Template shape}       &3.5 &3.3 &6.4 &8.7  \\      
   {Sample size}	  &1.5 &2.6 &4.1 &6.0  \\     
   {Total uncertainty}    &4.1 &4.3 &7.9 &10.7  \\      
   \hline \hline
 \end{tabular}
 \label{fbbsyst}
\end{table}

\section{\label{sec:Bckg} Background}
Three sources of background events are considered, ordered by relevance: events with at least one light-quark jet, $c\bar{c}$ events, and events with a misidentified muon. 

The asymmetry contribution from events with at least one light jet mistagged as a $b$-jet is checked using data~\cite{dataDrivenTechnique}. The obtained asymmetries are consistent with zero. 

The $c\bar{c}$ events are dominantly produced via gluon-gluon fusion. The fraction of $c\bar{c}$ pairs produced by $q\bar{q}$ annihilation that give rise to \AFB is smaller than in the \bbbar{} case. Therefore no asymmetry is assumed to arise from $c\bar{c}$ production. 

In events with misidentified muons no asymmetry is expected, as the distribution of the trajectories for misidentified muons is uniform as a function of detector solid angle. Therefore this background is also treated as symmetric.

A possible contribution to asymmetry can come from events where one of the jets is initiated by a $b$ quark, and the other by a $c$ quark; however, these events are produced via quark-gluon interactions, which is suppressed at Tevatron energies and therefore they are expected to contribute negligibly to the asymmetry~\cite{RodrigoKuhn, RodrigoKuhn2008}. 

As all sources of background are assumed to give negligible contribution to $b\bar{b}$ asymmetry we treat the background as symmetric. Nonetheless, we consider a possible asymmetry coming from background as a systematic uncertainty.

\section{\label{sec:unfold} Unfolding}
The measured signal distribution, $\vec{b}$, after background subtraction, defined as an eight-component vector of event frequencies corresponding to each of the histogram eight bins shown in Fig.~\ref{AfbInput}, is related to the underlying particle-level distribution, $\vec{x}$, by the relation

\begin{equation}
\label{eq:measVec}
\vec{b} = \mathrm{\bf{SA}}\vec{x}\;\mathrm{,}
\end{equation}

\noindent
where {\bf{A}} is a diagonal matrix that describes the acceptance in each bin of the measured distribution, and the non-diagonal smearing matrix {\bf{S}} describes the migration of events between bins due to the finite resolution of the CDF II detector and the reconstruction technique.

The binned data are multiplied by the inverse matrices to recover the true particle-level distribution from the background-subtracted one,

\begin{equation}
\label{eq:trueVec}
\vec{x} = \mathrm{\bf{A}}^{-1}\mathrm{\bf{S}}^{-1}\vec{b}\;\mathrm{.}
\end{equation}

Before applying the acceptance correction, we first remove the smearing effects of the resolution. To unfold the distribution using the {\bf{S}} matrix, an algorithm based on the singular-value decomposition (SVD) method is used~\cite{unfolding}. The SVD algorithm decomposes the non-diagonal smearing matrix $\mathrm{\bf{S}}$ as 
\begin{equation}
\label{eq:svd}
\mathrm{\bf{S}}=\mathrm{\bf{U\Sigma V}}^T\;\mathrm{,}
\end{equation}
\noindent
where {\bf{U}} and {\bf{V}} are orthogonal matrices ($\mathrm{\bf{U}}^T\mathrm{\bf{U}}=\mathrm{\bf{V}}^T\mathrm{\bf{V}}=1$), and $\mathrm{\bf{\Sigma}}={\text{diag}}\{s_{1},s_{2},...,s_{n}\}$ is a diagonal matrix of rank $n$ containing only non-negative entries, called singular values of {\bf{S}}, in decreasing order. The unfolding procedure is then reduced to the inversion of the diagonal matrix $\mathrm{\bf{\Sigma}}$. To avoid amplifying fluctuations caused by small elements of $\mathrm{\bf{\Sigma}}$, we use an {\em a priori} chosen regularization condition, {\bf{C}},  defined in Eq.~(\ref{CurvM}), which maximizes the ``smoothness" of the unfolded distribution by minimizing the second derivative~\cite{unfolding}. Hence, the expression we minimize to obtain the unfolded distribution $\vec{x'}$, which approximates the true distribution $\vec{x}$, is
\newline
\begin{center}
$(\mathrm{\bf{S}}\vec{{x'}}-\vec{b})^{T}\mathrm{\bf{B}}^{-1}(\mathrm{\bf{S}}\vec{x'}-\vec{b}) + \tau(\mathrm{\bf{C}}\vec{x'})^{T}(\mathrm{\bf{C}}\vec{x'})\;\mathrm{,}
$
\end{center}
\noindent
where {\bf{B}} is the covariance matrix of $\vec{b}$ and $\tau$ is the optimal regularization strength, which is related to the singular value of the smearing matrix {\bf{S}}. As the singular values $s_i$ are listed by decreasing absolute value, some index $k$, called the regularization parameter, exists such that the optimal $\tau$ is given by $\tau = s_{k}^{2}$. 

\begin{figure}[ht]
\includegraphics[width=\linewidth{}]{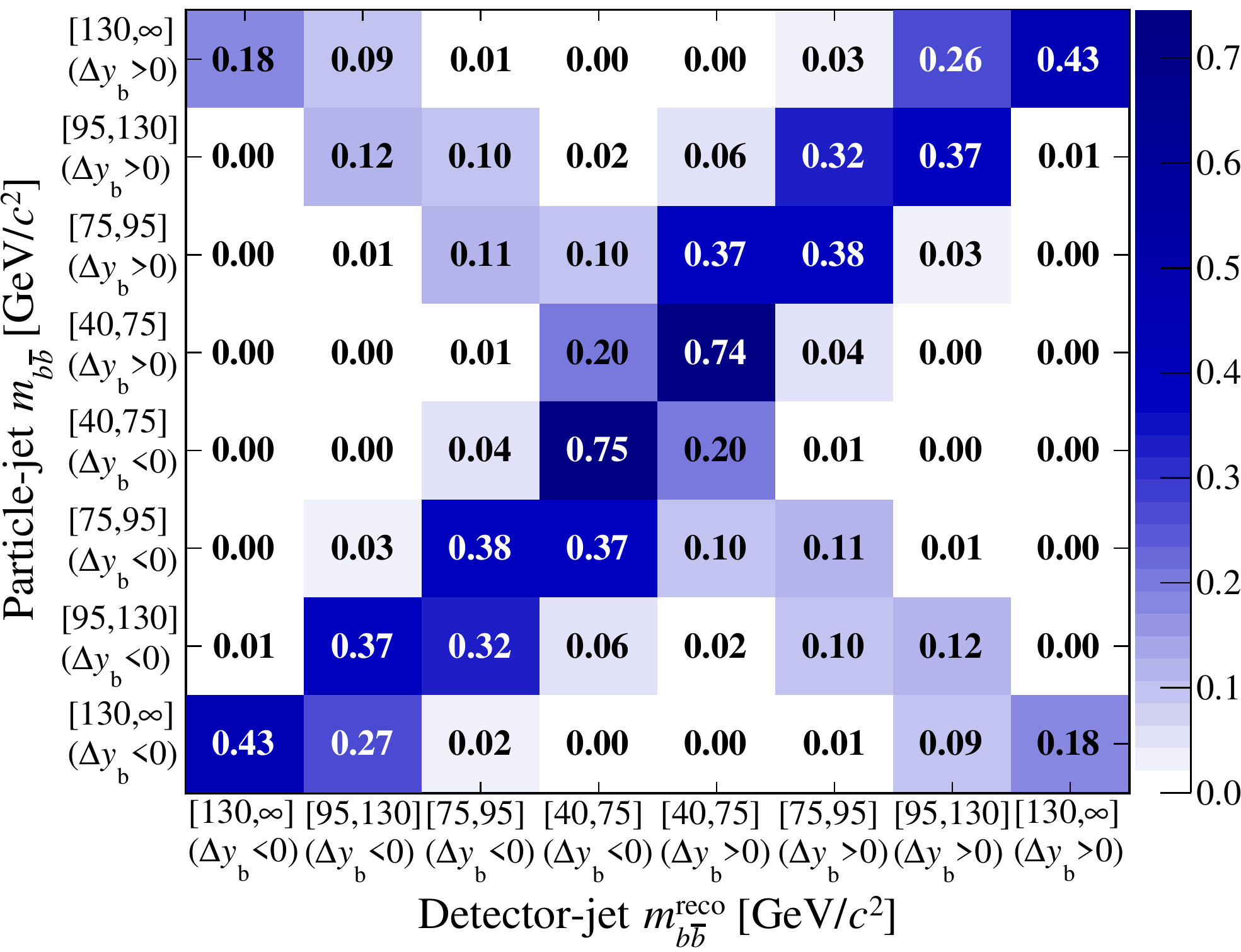}
\caption{Smearing matrix determined using simulated \bbbar events and used in the unfolding of the final results. Shading illustrates diagonalness of the smearing matrix and represents the level of probability that a given $m_{b\bar{b}}^{\rm reco}$ bin will be projected into a given \mbb bin.}
\label{RespM}}{%
\end{figure}

The smearing and acceptance matrices are modeled using the {\sc{pythia}} Monte Carlo sample together with the model of $Z/\gamma^{*}$ production of $b\bar{b}$ events.
Figure~\ref{RespM} shows the smearing matrix, which expresses the probability of observing a mass $m_{b\bar{b}}^{\rm reco}$ for a $b\bar{b}$ pair produced originally with mass \mbb{}.

\begin{figure}[ht]
\includegraphics[width=\linewidth{}]{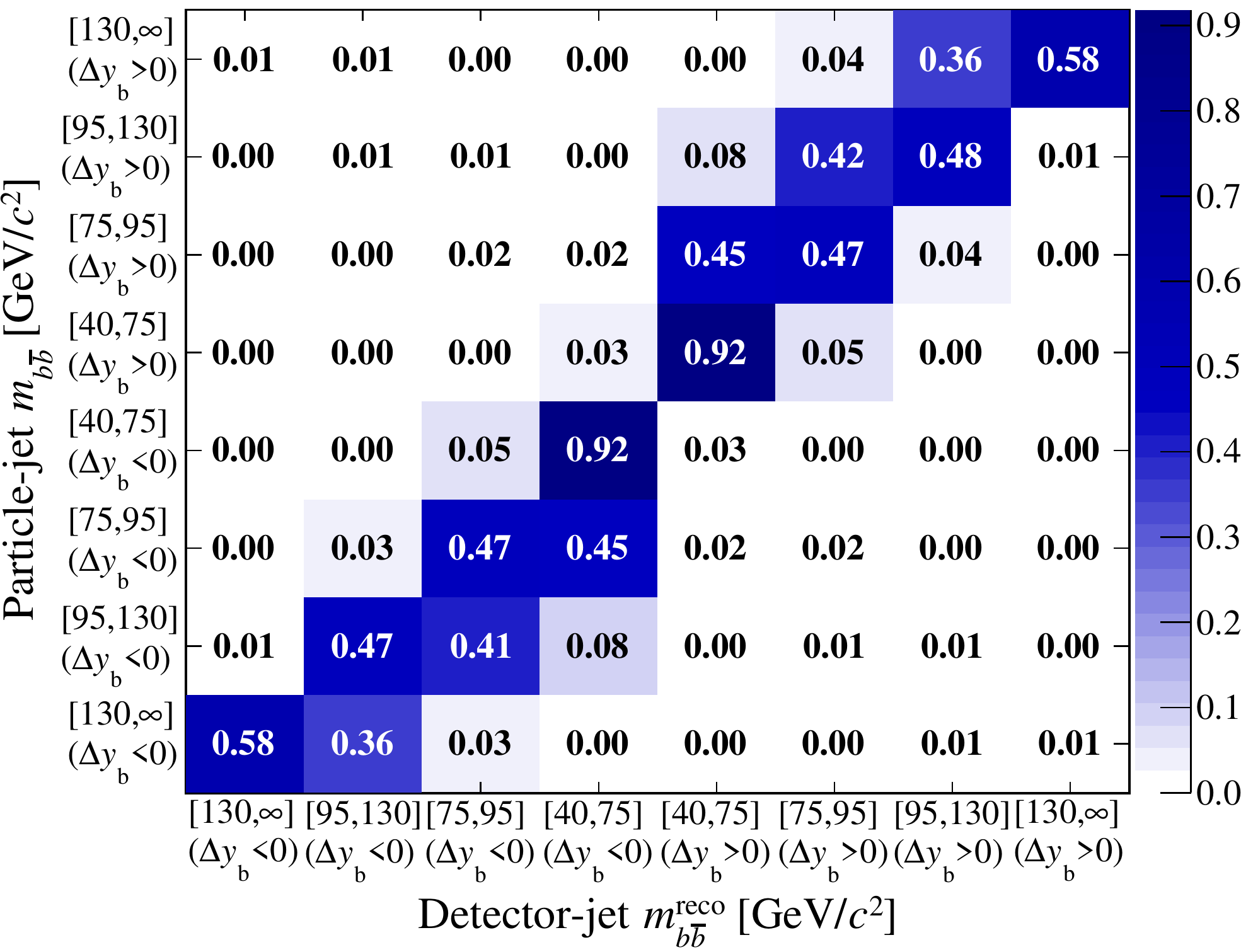}
\caption{Smearing matrix determined using simulated \bbbar events where no cascade decays and no $B^0_{(s)}-\overline{B}_{(s)}^0$ mixing occurs and used as a consistency check. Shading illustrates diagonalness of the smearing matrix and represents the level of probability that a given $m_{b\bar{b}}^{\rm reco}$ bin will be projected into a given \mbb bin.}
\label{pureRespM}}{%
\end{figure}

The smearing matrix is mostly diagonal with small contributions in the antidiagonal terms. We hypothesize that the antidiagonal terms originate from events where the sign of the muon charge changes due to $B^0_{(s)}-\overline{B}_{(s)}^0$ mixing or cascade $b\rightarrow c\rightarrow\mu$ decays. In both  these cases the sign of  muon charge is opposite to that of decaying $b$ quark, i.e., the muon incorrectly determines the sign of $b$ ($\bar{b}$) quark. To support this hypothesis, we present in Fig.~\ref{pureRespM} the smearing matrix for events where no $B^0_{(s)}-\overline{B}_{(s)}^0$ mixing or cascade $b\rightarrow c\rightarrow\mu$ decays occurred. 

In the unfolded distribution, the \mbb bins $[40;\,70]\,$\gevMns{} and $[130;\,\infty]\,$\gevMns{} are considered as the ``edge" bins. Therefore the corresponding elements of {\bf{C}} are $-1$ rather than $-2$: 
\begin{equation}
 C = \begin{pmatrix}
-1 & 1 & 0 & 0 & 0 & 0 & 0 & 0  \\
1 & -2 & 1 & 0 & 0 & 0 & 0 & 0  \\
0 & 1 & -2 & 1 & 0 & 0 & 0 & 0  \\
0 & 0 & 1 & -1 & 0 & 0 & 0 & 0  \\
0 & 0 & 0 & 0 & -1 & 1 & 0 & 0  \\
0 & 0 & 0 & 0 & 1 & -2 & 1 & 0  \\
0 & 0 & 0 & 0 & 0 & 1 & -2 & 1  \\
0 & 0 & 0 & 0 & 0 & 0 & 1 & -1  \\
\end{pmatrix}\;\mathrm{.}
\label{CurvM}
\end{equation}

We use the unfolding algorithm {\sc{roounfold}}~\cite{RooUndold} with the regularization parameter $k$ as input. For large $k$, close to the rank of the $\mathrm{\bf{\Sigma}}$ matrix, SVD unfolding is equivalent to pure matrix inversion. For small $k$ ($k \approx 1$), the regularization condition is strongly enforced.
To determine the best value of the regularization parameter $k$, we introduce an asymmetry into the {\sc{pythia}} Monte Carlo sample by reweighting the selected events. For each \mbb bin, several asymmetry variations around the predicted values are tested by running $1000$ pseudoexperiments for each regularization parameter $k$. After unfolding, the measured asymmetry is compared with the generated asymmetry. Minimizing the bias from the comparison, $k=4$ was chosen as the optimal regularization parameter.

\section{\label{sec:Syst} Systematic uncertainties}

The sources of systematic uncertainties considered in this analysis come from modeling of the geometric and kinematic acceptance, estimation of the background, and possible asymmetry of the background. Modeling of the geometric and kinematic acceptance includes effects of initial- and final-state radiation (ISR and FSR), and jet-energy scale (JES). These are estimated by varying ISR, FSR, and the JES in the simulation. The uncertainty due to the amount of background is estimated by varying the $b\bar{b}$ fraction within its uncertainty. The systematic uncertainty due to a possible asymmetry of the background is estimated by simulating $\pm1\%$ asymmetries in the background distributions. The total systematic uncertainties for different \mbb{} bins are summarized in Table~\ref{systSummary}.

\begingroup
\squeezetable
\begin{table}[ht]
 \caption{Absolute systematic uncertainties of $\AFB$ in percentage.}
 \begin{tabular}{lS[table-format=2.1]S[table-format=2.1]S[table-format=2.1]S[table-format=2.1]S[table-format=2.1]}
   \toprule
   \hline \hline
   \multicolumn{6}{c}{Systematic uncertainty of \afbb$[\si{\%}]$} \\
   \multirow{2}{*}{} & \multicolumn{4}{c}{\mbb [\gevMns{}]} & \multirow{2}{*}{  Integrated} \\
   & {40~--~75 } & { 75~--~95 } & { 95~--~130 } & {$> 130$} & \\
   \hline
   {ISR/FSR}                 & 0.09 & 0.07 & 0.06 & 0.12 & 0.05 \\
   {JES}                     & 0.24 & 0.15 & 0.02 & 0.10 & 0.10 \\
   {$f_{b\bar{b}}$ uncert.}  & 0.06 & 0.06 & 0.04 & 0.01 & 0.04 \\
   {Background \AFB{}}       & 0.11 & 0.17 & 0.27 & 0.34 & 0.17 \\
   {Total}                   & 0.29 & 0.24 & 0.28 & 0.37 & 0.22 \\
   \hline \hline
 \end{tabular}
\label{systSummary}
\end{table}
\endgroup

\section{\label{sec:Res}Results}
In this analysis we measure three asymmetries, the raw observed asymmetry, which include effects from background asymmetries, and detector acceptance and resolution; the background-subtracted raw $b\bar{b}$ asymmetry, which is corrected for asymmetries induced by backgrounds but not for effects due to detector acceptance and resolution; and the particle-level asymmetry, which is corrected for background and detector effects. The first two asymmetries are shown to demonstrate the effect of the performed corrections. The results are summarized in Table~\ref{AfbTab}, where the \AFB{} dependence on \mbb{} and the integrated asymmetry are presented. The final results, with statistical and systematic uncertainties added in quadrature, are summarized in the last column. The significant difference between the raw \AFB{} and the particle level \AFB{} comes from the interplay of the unfolding procedure and the small number of events in the highest \mbb bin.

\begin{table}[!ht]
 \caption{Results of the \AFB{} measurements as functions of $b\bar{b}$ invariant mass. The integral values for each bin are shown.}
 \begin{tabular}{cccc}
 \hline \hline
  & \multicolumn{3}{c}{\AFB $[\si{\%}]$} \\
 \multirow{3}{*}{\mbb [\gevMns{}]} &  {Observed}	& {Background}	& {Particle} \\
				   &  {raw asymmetry}	& {subtracted}	& {level} \\
 				   & \multicolumn{2}{c}{(statistical uncertainty only)}		& {(stat+syst)} \\
 \hline
 {40 -- 75}  		 & $0.47 \pm 0.49$ & $0.50 \pm 0.54$  & $0.83 \pm 0.83$  \\
 {75 -- 95}  		 & $0.55 \pm 0.61$ & $0.60 \pm 0.70$  & $1.54 \pm 0.73$  \\
 {95 -- 130}    	 & $0.70 \pm 0.71$ & $0.83 \pm 0.90$  & $0.92 \pm 0.87$  \\
 {$> 130$}	 	 & $0.32 \pm 0.91$ & $0.43 \pm 1.33$  & $2.08 \pm 1.10$  \\
 Integrated		 & $0.52 \pm 0.32$ & $0.58 \pm 0.37$  & $1.17 \pm 0.71$  \\
 \hline \hline
 \end{tabular}
 \label{AfbTab}
\end{table}
\begin{figure}[ht!]
\includegraphics[width=\linewidth{}]{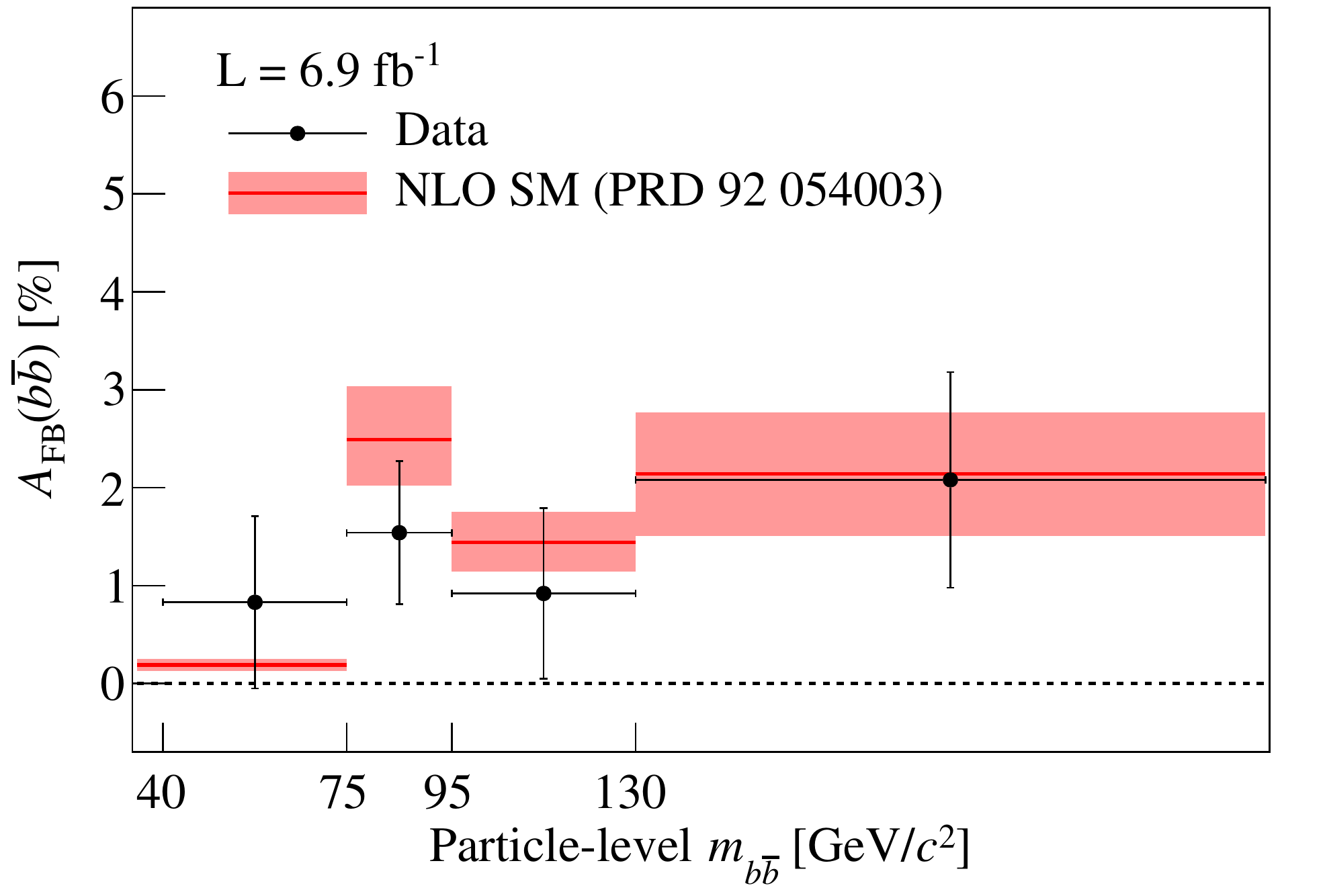}
\caption{Measured \afbb{} as a function of particle-level \mbb{}. The data are compared with the NLO theoretical prediction~\cite{Murphy}.}
\label{AfbFinal}
\end{figure}

Figure~\ref{AfbFinal} shows the comparison of the results with the theoretical prediction~\cite{Murphy}, which is calculated at the parton level using a different lower threshold for the lowest \mbb{} bin. However, the parton-to-particle corrections of the theory predictions are expected to be small compared to the experimental uncertainties. An indication of an increase as a function of \mbb{} is visible in $\afbb{}$ measured at particle level, with an enhancement around the $Z$-pole mass similar to that predicted theoretically. The measured integrated asymmetry of $(1.2 \pm 0.7)\%$ is consistent with the SM prediction. The measurement is compared with the SM and non-SM predictions in Ref.~\cite{Murphy}, where the models with a \gevM{100} axigluon are disfavored.

\section{Conclusions}
We present a measurement of the forward-backward asymmetry in the production of $b\bar{b}$ pairs in proton-antiproton collisions at $\sqrt{s} = 1.96$ TeV using a sample from 6.9 fb$^{-1}$ of integrated luminosity collected by the CDF II detector. The measured value of the integrated asymmetry is $A_\mathrm{FB}(b\bar{b}) = (1.2 \pm 0.7)$\%. In addition, the dependence of the asymmetry $A_\mathrm{FB}$ on $b\bar{b}$-pair invariant mass, \mbb{}, is measured. A tendency to an increase of \afbb with \mbb{} invariant mass is observed. The experimental value of \afbb around the $Z$ boson mass follows the theoretical expectation that predicts a local increase due to the contribution of electroweak processes~\cite{Murphy}. This results is consistent with SM expectations and extends previous findings \cite{afbbb} to \mbb{} values down to \gevM{40}.

\section*{Acknowledgements}
We thank the Fermilab staff and the technical staffs of the participating institutions for their vital contributions. This work was supported by the U.S. Department of Energy and National Science Foundation; the Italian Istituto Nazionale di Fisica Nucleare; the Ministry of Education, Culture, Sports, Science and Technology of Japan; the Natural Sciences and Engineering Research Council of Canada; the National Science Council of the Republic of China; the Swiss National Science Foundation; the A.P. Sloan Foundation; the Bundesministerium f\"ur Bildung und Forschung, Germany; the Korean World Class University Program, the National Research Foundation of Korea; the Science and Technology Facilities Council and the Royal Society, UK; the Russian Foundation for Basic Research; the Ministerio de Ciencia e Innovaci\'{o}n, and Programa Consolider-Ingenio 2010, Spain; the Slovak R\&D Agency; the Academy of Finland; the Australian Research Council (ARC); and the EU community Marie Curie Fellowship contract 302103.

\bibliography{Afbb}

\begin{thebibliography}{45}
\bibitem [{Jet()}]{JetDef}%
  \BibitemOpen
  \href@noop {} {}\bibinfo {note} {By jet we understand an assembly of
  associated narrowly-clustered particles.}\BibitemShut {Stop}%
\bibitem [{Coo()}]{CoordinateSys}%
  \BibitemOpen
  \href@noop {} {}\bibinfo {note} {A cylindrical coordinate system with the $z$
  axis directed along the proton beam is used. The polar angle $\theta$ is
  defined with respect to the proton beam direction and $\phi$ is the azimuthal
  angle about the $z$ axis. The component of the particle momentum transverse
  to the beam line is $p_{T} = p \sin\theta$. The pseudorapidity is defined as
  $\eta = -\ln \tan(\theta/2)$. Jets are reconstructed by a
  calorimeter-tower-clustering cone algorithm with a cone size of $\Delta R
  \equiv \sqrt{(\Delta{\eta})^2 + (\Delta{\phi})^2} = 0.4$. The transverse
  energy, $E_\mathrm{T}$, is defined to be $E\sin\theta$, where $E$ is the
  reconstructed jet energy.}\BibitemShut {Stop}%
\bibitem [{rap()}]{rapidity}%
  \BibitemOpen
  \href@noop {} {}\bibinfo {note} {The rapidity is defined as $y = (1/2) \ln (E
  + p_z / E - p_z)$, where $E$ and $p_z$ are particle total energy and
  longitudinal momentum, respectively.}\BibitemShut {Stop}%
\end{thebibliography}
\bibliographystyle{apsrev4-1-JHEPfix}

\end{document}